\documentclass[preprint,aps]{revtex4}

\newcommand {\bea}{\begin{eqnarray}}
\newcommand {\eea}{\end{eqnarray}}
\newcommand {\be}{\begin{equation}}
\newcommand {\ee}{\end{equation}}

\usepackage{graphicx}

\begin{document}


\title{Euclidean Correlation Functions in a\\
       Holographic Model of QCD}

\author{T.~Sch\"afer}

\affiliation{Department of Physics, North Carolina State University,
Raleigh, NC 27695}

\begin{abstract}
We compute euclidean coordinate space correlation functions in a 
holographic model of QCD. We concentrate, in particular, on channels 
that are related to the $U(1)_A$ problem, the flavor-singlet 
axialvector, pseudoscalar meson, and pseudoscalar glueball 
(topological charge) correlator. We find that even a very simple
holographic model defined on a slice of $AdS_5$ provides a 
qualitatively correct description of QCD correlation functions. 
We study the role of anomaly terms, and show that both euclidean 
positivity and low energy theorems based on the axial anomaly 
relation are correctly implemented.  We compare the results with 
expectations from an instanton model of the QCD vacuum.

\end{abstract}

\maketitle
\section{Introduction}
\label{sec_intro}

 Recently, much progress has been achieved in realizing the old
idea that large $N$ gauge theories are related to string theory
\cite{'t Hooft:1973jz,Migdal:1977nu}. The crucial development in 
this regard was the discovery of the correspondence between ${\cal N}=4$ 
super Yang-Mills theory and type IIB string theory on $AdS_5\times S_5$ 
\cite{Maldacena:1997re}. Current effort is aimed at finding string 
duals of theories more closely related to QCD, in particular theories that 
exhibit asymptotic freedom, confinement and chiral symmetry breaking. Two 
avenues of research are being pursued. The top-down approach is based on 
the study of generalizations of the original AdS/CFT setup that incorporate 
fundamental fermions, chiral symmetry breaking and confinement 
\cite{Polchinski:2000uf,Karch:2002sh,Babington:2003vm,Sakai:2004cn}.
The bottom-up approach is founded on models of QCD that are defined 
on warped higher dimensional spaces and incorporate the general 
principles, in particular holography, of the AdS/CFT correspondence
\cite{Migdal:1977ut,DaRold:2005zs,Erlich:2005qh}. The aim of 
the bottom-up approach is to find model independent features of 
the holographic description, to provide guidance for the top-down
approach, and to develop a geometric language for thinking about 
gauge theories like QCD. 

 Most of the published works on holographic models focus on the 
hadronic spectrum or on hadronic form factors. In this work we shall 
study euclidean correlation functions in a simple holographic model 
of QCD. Euclidean correlators provide a bridge between perturbation
theory and the operator product expansion, which control the 
short distance behavior, and the mass gap, which governs the long
distance behavior \cite{Shifman:1978bx}. Phenomenological information 
on euclidean correlation functions comes from lattice data and from 
spectral functions measured in processes like $e^+e^-$ annihilation into
hadrons or hadronic $\tau$ decays \cite{Shuryak:1993kg,Schafer:2000rv}.

 We will focus, in particular, on correlation functions related to 
the $U(1)_A$ problem. This includes the flavor-singlet axialvector, 
pseudoscalar meson, and pseudoscalar glueball channel. The correlators 
in these channels have been studied in the instanton model
\cite{Geshkenbein:1979vb,Novikov:1979ux,Schafer:1994fd,Schafer:1996wv}, 
and there are some recent results from lattice simulations
\cite{BilsonThompson:2002jk,Horvath:2005cv,Moran:2007nc}. 
There are a number of qualitative questions related to the $U(1)_A$
sector in QCD that are still not fully understood. These include
the origin of the topological susceptibility in pure gauge QCD, the 
microscopic mechanism that causes the susceptibility to vanish in QCD 
with massless fermions, and the role of instantons or other topological
objects. The present work is motivated by the idea that holographic 
models of QCD can shed some light on these questions.

\section{Holographic model}
\label{sec_ss}

 We consider the holographic model introduced by Erlich et 
al.~\cite{Erlich:2005qh} and extended to the flavor singlet sector 
by Katz and Schwartz \cite{Katz:2007tf}. The model is defined by 
the 5-dimensional lagrangian 
\bea
\label{s_5d}
  S &=& \int d^5x\,\sqrt{g}\, 
  \Bigg\{ - \frac{1}{4 g_5^2} 
    {\rm Tr} \left( F_L^2 + F_R^2 \right) +
    {\rm Tr} \left( |D X|^2 + 3 |X|^2 \right) 
    \\
    & & \mbox{}\hspace{2.25cm}
     + \frac{1}{2}|D Y|^2  
     + \frac{\kappa_0}{2}\left( Y^{N_f} \det(X)  +  h.c. \right) 
       \Bigg\} \nonumber 
\eea
where $X=X^at^a$ is a scalar field, $Y$ is a complex flavor singlet 
scalar, $F_{\mu\nu} = \partial_\mu A_\nu -\partial_\nu A_\mu - i[A_\mu, 
A_\nu]$ (for L/R) is the field strength tensor corresponding to the 
gauge field $A_{L,R}=A_{L,R}^at^a$. Here, $t^a$ are the generators 
of $U(3)_F$. The covariant derivative is $D_\mu X=\partial_\mu X - 
iA_{L\mu} X + iX A_{R\mu}$. The model is defined on an $AdS_5$ metric
\be
ds^2 =  \frac{1}{z^2}
  \left(-dz^2 + dx^\mu dx_\mu\right),
\ee
with a ``hard wall'' cutoff $z_m$ ($0\leq z\leq z_m$). The 5-dimensional 
masses of the fields are determined by the correspondence between fields 
on $AdS_5$ and operators on the boundary, $m_5^2=(\Delta-p)(\Delta+p-4)$. 
Here, $m_5$ is the 5-d mass, and $\Delta$ is the dimension of a $p$-form 
operator on the boundary. The scalar field $X$ corresponds to the 
operator $\bar{q}_iq_j$ with $\Delta=3$ and the complex scalar field
$Y$ corresponds to $g^2(GG+i\tilde{G}G)$ with $\Delta=4$. The gauge 
field $A_\mu$ has $p=1$ and couples to the $\Delta=3$ operator 
$\bar{q}_i\gamma_\mu q_j$. The terms in the action that only 
involve $F_{L,R}$ and $X$ preserve the full $U(3)\times U(3)$ 
symmetry \cite{Erlich:2005qh}, and the last term generates
the anomalous Ward identities of QCD \cite{Katz:2007tf}.

 The expectation values of $X$ and $Y$ are determined by classical 
solutions to the equations of motion in the presence of sources $M$ 
and $c$, where $M$ is the quark mass matrix, and $c$ is related to 
the strong coupling constant (see equ.~(\ref{c_match}) below). We have 
\be
\label{XY_vev}
  \langle X_{ij} \rangle = \sigma_{ij} z^3 + M_{ij} z, 
  \hspace{1cm}
  \langle Y \rangle       = \Xi\, z^4 + c ,
\ee
where $\sigma_{ij}$ and $\Xi$ correspond to the vacuum expectation 
values $\langle\bar{q}_i q_j \rangle$ and $\langle G^2 \rangle$ 
of operators conjugate to the sources $M$ and $c$. In a more complete, 
top-down, model the expectation values $\sigma$ and $\Xi$ are determined 
dynamically \cite{Babington:2003vm,Sakai:2004cn}. Alternatively, one may 
incorporate the dynamics of chiral symmetry breaking into a boundary action 
for $X$ and $Y$ \cite{DaRold:2005zs}. Here, rather than specify these 
boundary terms, we shall take $\sigma$ and $\Xi$ to be free parameters 
of the model. In this work we are not specifically interested in flavor 
mixing (see \cite{Katz:2007tf}) and we will set $\sigma_{ij}=\delta_{ij}
\sigma$ and $M_{ij}=\delta_{ij}m$. 

\begin{figure}[t]
\begin{center}
\includegraphics[width=11cm]{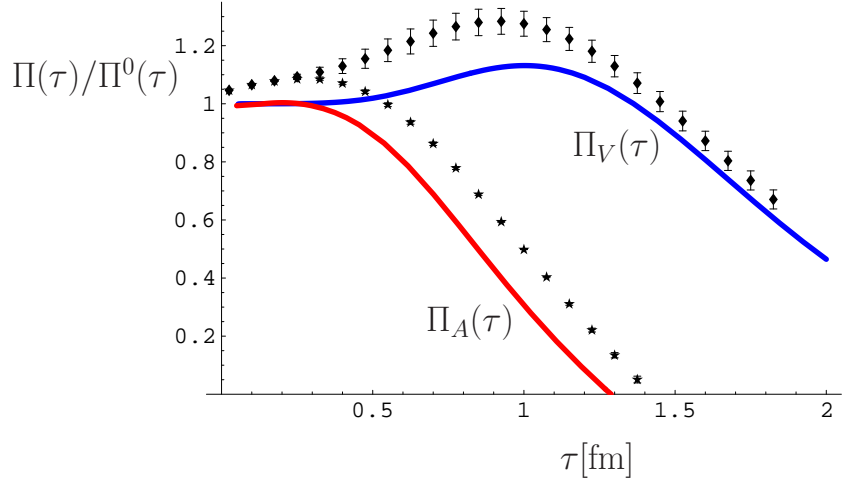}
\end{center}
\caption{\label{fig_piv}
Vector and axialvector current correlation functions. We show the
ratio of the correlation function to the free correlator as a 
function of euclidean separation $\tau$. The solid curves show the
result in the holographic model. The data points are taken from an 
analysis of Aleph data on hadronic tau decays \cite{Schafer:2000rv}.}
\end{figure}

 The coupling constants $g_5$ and $c$ are determined by matching 
the short distance behavior of correlation functions to QCD.
Consider the vector current correlation function
\be 
\Pi_V(Q^2)\delta^{ab}\left(q_\mu q_\nu -g_{\mu\nu}q^2\right)
 = \int d^4x\, e^{iqx} \, 
   \langle J_\mu^a(x)J_\nu^b(0) \rangle ,
\ee
where $J_\mu^a=\bar{q}\gamma_\mu t^aq$ and $Q^2=-q^2$. Using the 
AdS/CFT dictionary we get
\be
 \Pi_{V}(Q^2) = -\frac{1}{g_5^2Q^2}
 \left. \frac{V(q,z)\partial_z V(q,z)}{z}\right|_{z=\epsilon} \, , 
\ee
where $V(q,z)$ is the bulk-to-boundary propagator, $V_\mu(q,z)=
V(q,z)V^0_\mu(z)$, and $V^0_\mu(q)$ is the boundary value of the 
field. The propagator has to satisfy the equation of motion. In 
the $V_z(x,z)=0$ gauge the linearized equation of motion for the 
gauge field is
\be
\label{eom_v}
  \partial_z\left(\frac{1}{z} \partial_z V_\mu^a(q,z) \right) 
    + \frac{q^2}{z} V_\mu^a(q,z) = 0.
\ee
A solution on the complete $AdS_5$ space with $V(q,z\to 0)=1$ is 
given by $V(q,z)=-\frac{\pi}{2}(qz) Y_1(qz)$ and
\be
  \Pi_{\rm V}(Q^2) = -\frac{1}{2g_5^2} \log \left(Q^2\right).
\ee
This result can be compared to the perturbative one-loop contribution 
$\Pi_V(Q^2)=-N_c/(24\pi^2)\cdot \log(Q^2)$. This leads to the 
matching condition
\be
\label{g5_match}
g_5^2 = \frac{12\pi^2}{N_c} .
\ee
We observe that $g_5^2\sim 1/N_c$ which shows that vector mesons at 
large $N_c$ are weakly coupled, as expected from general large $N_c$
arguments. The constant $c$ is fixed by matching the flavor-singlet
axialvector correlation function and using the axial anomaly relation.  
The result is \cite{Katz:2007tf}
\be 
\label{c_match}
 c = \sqrt{2N_f}\, \frac{\alpha_s}{2\pi^2},
\ee
and we shall follow \cite{Katz:2007tf} and allow $\alpha_s$ to 
run as a function of $z$ according to the one-loop beta function. 
Having fixed the coupling constants we can now compute the correlation 
functions. For this purpose we shall employ the representation of the 
Green function in terms of eigenfunctions of the five-dimensional 
Sturm-Liouville problem. In the vector channel
\be 
\label{piv_mode}
\Pi_V(Q^2) = \sum_k \frac{f_{\rho,k}^2}
    {m_{\rho,k}^2\left(Q^2+ m_{\rho,k}^2\right)}
\ee
where $q^2=m_{\rho,k}^2$ is an eigenvalue of equ.~(\ref{eom_v})
subject to the boundary conditions $V(0)=0$ and $V'(z_m)=0$.
The eigenfunctions are normalized according to  
\be 
 \int dz\, \frac{1}{z}\, V_k(z)V_l(z) = \delta_{kl},
\ee
and the decay constants $f_{\rho,k}$ are given by 
\be 
 f_{\rho,k}= \frac{1}{g_5} \left. \frac{V'(z)}{z}\right|_{z=\epsilon}.
\ee
The eigenmode representation (\ref{piv_mode}) also determines the 
spectral function and the euclidean correlator. The spectral function
is a sum of delta-functions, and the euclidean correlation  function is
\be 
\Pi_V(\tau) = \sum_k f_{\rho,k}^2 D(m_{\rho,k},\tau),
\hspace{1.5cm}
D(m,\tau)=\frac{m}{4\pi^2\tau}\, K_1(m\tau),
\ee
where $D(m,\tau)$ is the euclidean propagator of a scalar particle 
with mass $m$. In the vector channel the spectrum can be obtained
almost in closed form. The eigenvalues $m_{\rho,k}$ are determined
by the zeros of the Bessel function of the first kind and order 
zero, $J_0(m_{\rho,k}z_m)=0$. Using the location of the first zero 
and the physical mass of the rho meson we can set the scale, 
$z_m=0.62$ fm. The coupling constant of the k'th resonance is 
given by 
\be 
 f_{\rho,k}= \frac{\sqrt{2}}{g_5}\,
  \frac{m_{\rho,k}}{z_m J_1(m_{\rho,k}z_m)}.
\ee
The euclidean correlation function is shown in Fig.~\ref{fig_piv}.
We plot the ratio of the euclidean correlator $\Pi_V(\tau)$ over 
the correlator in the non-interacting theory, $\Pi_V^0(\tau) = 
N_c/(3\pi^4\tau^6)$. Because of asymptotic freedom this ratio 
approaches one as $\tau\to 0$. The result can be compared to the 
experimental results compiled in \cite{Schafer:2000rv}. The 
agreement is quite good, but at short distances perturbative
logarithms and power corrections are missing. The short distance
behavior can be improved by adding extra terms to the action given 
in equ.~(\ref{s_5d}), but this is beyond the scope of the present 
work. The agreement with experimental data is comparable to 
resonance saturation models constrained by Weinberg sum rules
and other QCD inputs \cite{Shifman:2000jv,Cata:2005zj}. This is 
not surprising, as equ.~(\ref{piv_mode}) is equivalent to the 
resonance saturation ansatz, equ.~(\ref{g5_match}) is the 
statement of local duality, and Weinberg sum rules are realized
as a consequence of chiral symmetry. 

\section{Pseudoscalar and Axialvector Correlation Functions}
\label{sec_ps}

\begin{figure}[t]
\begin{center}
\includegraphics[width=11cm]{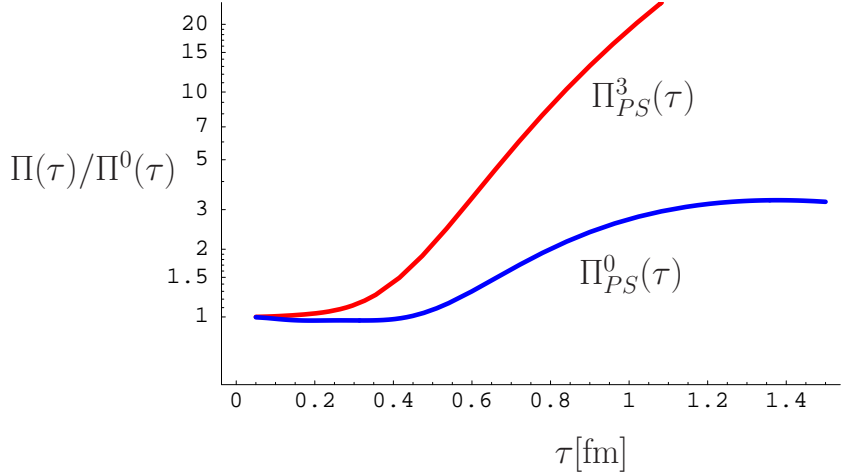}
\end{center}
\caption{\label{fig_pip}
Pseudoscalar octet and singlet current correlation functions. The 
correlation functions are normalized to the correlator in the free
theory. }
\end{figure}

 The equations of motion for the transverse components of the 
axialvector gauge field is
\be
   \partial_z\left(\frac{1}{z} \partial_z A^a_\mu \right) 
  + \frac{q^2}{z} A^a_\mu
- \frac{g_5^2 v^2}{z^3} A^a_\mu =0 ,
\ee
where we have introduced the notation $v=mz+\sigma z^3$ for the vacuum 
expectation value of the scalar field. This equation differs from the 
equation for the vector field by an extra contribution due to (explicit
and spontaneous) chiral symmetry breaking. The longitudinal component 
of the axialvector current mixes with the pseudoscalar current, and 
pseudoscalar fluctuations have be to taken into account. We define octet 
$\eta^a$ ($a=1,\ldots, 8$) and singlet $\eta^0,a$ fields by
\be
X_{ij} = \langle X_{ij} \rangle \exp(i\eta^a t^a),
\hspace{1cm}
Y = \langle Y \rangle \exp(ia).
\ee
We also define the longitudinal component of the axialvector field 
$\partial_\mu\varphi^a=A^a_\mu-A^a_{\mu\perp}$ (in $A^a_z=0$ gauge). 
We first consider the flavor non-singlet sector. The equations of 
motion for the longitudinal axialvector $\varphi^a$ and the pseudoscalar 
field $\pi^a$ are
\bea
\label{oct_eom_1}
  \partial_z\left(\frac{1}{z} \partial_z \varphi^a \right) 
+\frac{g_5^2 v^2}{z^3} (\pi^a-\varphi^a) &=& 0 \, .  \\
\label{oct_eom_2}
  -q^2\partial_z\varphi^a
   + \frac{g_5^2 v^2}{z^2} \partial_z \pi^a &=&0 \, . 
\eea
The solutions are normalized according to
\be 
\int dz\, \frac{v^2}{z^3} \, \pi^a_k
  \left(\pi^a_l-\varphi^a_l \right) = \delta_{kl} 
\ee
and the pion decay constant is given by 
\be
f_{\pi,k} = \frac{1}{g_5^2}
   \left. \frac{\partial_z\varphi^a_k}{z}\right|_{\epsilon} .
\ee
We consider $N_c=3$ and set $m=2.23$ MeV and $\sigma=(323\,{\rm MeV})^3$ 
in order to reproduce the experimental values of $m_\pi$ and $f_\pi$ for the 
lowest pion excitation. The trace of the axialvector correlation function, 
$\Pi_A \equiv \Pi_{A\,\mu}^{\;\;\mu}$, is shown in Fig.~\ref{fig_piv}. The 
splitting between the vector and axialvector correlators is related to the 
difference in mass and coupling between the vector and axialvector mesons,
and to the pion contribution. The latter dominates at large distance and
causes $\Pi_A$ to become negative. In the model considered here chiral 
symmetry is restored high in the spectrum ($m_{\rho,k}^2\to m_{a_1,k}^2$ 
as $k\to \infty$) and the vector and axialvector correlators are very 
nearly degenerate for $\tau<0.5$ fm. 

In the singlet sector there is also mixing with the pseudoscalar 
glueball field $a$. The equations of motion are
\bea 
\label{sing_eom_1}
\partial_z\left( \frac{1}{z} \partial_z\varphi^0\right) 
 -g_5^2 \frac{v^2}{z^3} \left( \varphi^0-\eta^0 \right) 
 -g_5^2 \frac{c^2}{z^3} \left( \varphi^0- a\right) &=& 0\, , \\
\label{sing_eom_2}
\partial_z\left( \frac{c^2}{z^3} \partial_z a\right) 
 + q^2  \frac{c^2}{z^3} \left( a - \varphi^0\right)
 + \kappa \frac{v^{N_f}}{z^5} \left( \eta^0 -a \right) &=& 0\, , \\[0.2cm]
\label{sing_eom_3}
q^2 z^2 \partial_z \varphi^0 -
g_5^2 v^2 \partial_z \eta^0  - 
g_5^2 c^2 \partial_z a  &=& 0 \, , 
\eea
where we have defined $\kappa=c^{N_f}\kappa_0$. The normalization 
condition is 
\be
\int dz\, \left[ \frac{v^2}{z^3} \eta^0_k\left(\eta^0_l-\varphi^0_l\right)
 + \frac{c^2}{z^3} a_k\left( a_l- \varphi^0_l \right) \right] = 
 \delta_{kl} \, . 
\ee
For a normalized eigenmode the coupling constants to the 
axialvector, pseudoscalar quark-anti-quark, and pseudoscalar 
glueball current are given by 
\be
 f_{\eta',k} = \frac{1}{g_5^2}
   \left. \frac{\partial_z\varphi^0_k}{z}\right|_{\epsilon}\, ,
\hspace{0.75cm}
 \lambda_{\eta',k} = m
   \left. \frac{\partial_z\eta^0_k}{z}\right|_{\epsilon}\, ,
\hspace{0.75cm}
 h_{\eta',k} = \frac{c^2}{\sqrt{2N_f}}
   \left. \frac{\partial_z a_k}{z^3}\right|_{\epsilon}\, . 
\ee
In Fig.~\ref{fig_pip} we show the pseudoscalar correlation 
functions 
\be
 \Pi_P(Q^2)\delta^{ab} = \int d^4x\, e^{iqx} \langle 
  \bar{q}t^a\gamma_5q(x) \bar{q}t^b\gamma_5q(0)\rangle
\ee
in the flavor singlet and non-singlet sector. In the non-singlet 
sector the correlation function is dominated by the light pion 
excitation. In the singlet sector the would-be Goldstone boson 
receives an extra contribution to its mass from the anomaly term. 
We have set $\kappa=20$ and find $m_{\eta'}=650$ MeV. (For $\kappa
\gg1$ the mass of the $\eta'$ becomes weakly dependent on the value 
of $\kappa$. We also note that the experimental result for the $\eta'$ 
mass, $m_{\eta'}=957$ MeV, can be reproduced if the strange quark mass
is taken into account.) Like chiral symmetry, the axial $U(1)_A$ 
symmetry is effectively restored in the highly excited part of the 
spectrum. For $k\gg 1$ the equations of motion for $\eta^0,\varphi^0$ 
and $a$ effectively decouple, and the solutions to 
Eqs.~(\ref{sing_eom_1}-\ref{sing_eom_3}) alternate between
solutions to the equations of motion in the non-singlet and the 
glueball sector of the theory. 

\begin{figure}[t]
\begin{center}
\includegraphics[width=11cm]{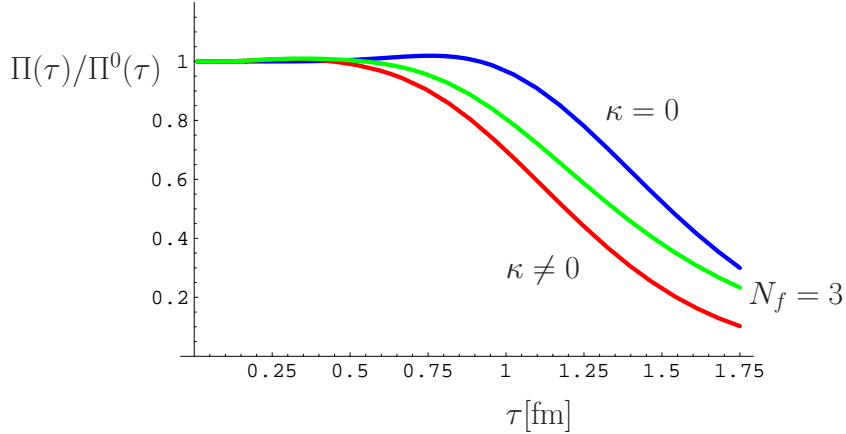}
\end{center}
\caption{\label{fig_pig}
Pseudoscalar glueball correlation functions. The correlation functions
are normalized to free field behavior. The curve labeled $\kappa=0$
corresponds to the pure gauge theory without anomaly term, whereas
the curve labeled $\kappa\neq 0$ includes the effects of an anomaly
term. The curve labeled $N_f=3$ shows the result in full QCD, 
including mixing with the $\eta'$. }
\end{figure}

\section{Topological charge correlator}
\label{sec_top}

\begin{figure}[t]
\begin{center}
\includegraphics[width=9cm]{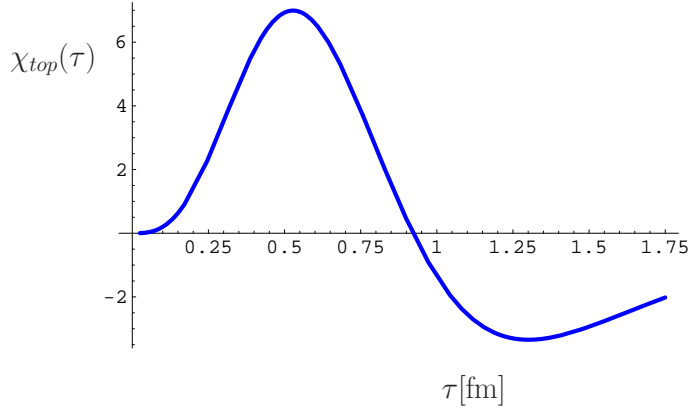}
\end{center}
\caption{\label{fig_top_x}
Integrand for topological susceptibility in pure gauge QCD without
an anomaly term. The integrand changes sign at $z\sim 1$ fm, and
the topological susceptibility vanishes. }
\end{figure}

 The pseudoscalar glueball correlation function is 
\be
\label{pi_top}
\Pi_P(Q^2) = -\int d^4x\, e^{iqx} \,
 \left\langle \frac{\alpha_s}{8\pi}G\tilde{G}(x) 
         \frac{\alpha_s}{8\pi}G\tilde{G}(0)\right\rangle .
\ee
Note that $\alpha_s/(8\pi)\cdot G\tilde{G}$ is the topological 
charge density. In order to study the topological charge correlator 
in the holographic model we begin with the equations of motion in 
the pure gauge sector of the theory. We have
\be 
\label{eom_a}
\partial_z\left( \frac{c^2}{z^3} \partial_z a\right) 
 + q^2  \frac{c^2}{z^3}  a 
 - \frac{\bar{\kappa}}{z^5} a = 0\, ,
\ee
where $\bar\kappa=\kappa v^{N_f}$. The solutions of this equation 
are particularly simple in the limit $c=const$ and $\bar{\kappa}
\to 0$. In this case the eigenvalues in the pseudoscalar glueball
channel are given by the solutions of $J_1(m_{G,k}z_m)=0$ and
the coupling constants are 
\be  
 h_{G,k}= \frac{c}{2\sqrt{N_f}}
  \frac{m_{G,k}^2}{z_m J_2(m_{G,k}z_m)}\, .
\ee
The groundstate in the pseudoscalar glueball channel is heavier than 
the ground state rho meson by a factor $\sim 1.5$. The glueball 
correlation function is shown in Fig.~\ref{fig_pig}. The shape of
the correlation function is similar to the vector meson correlator,
but since the mass scale is larger the correlator is smaller. 
The topological susceptibility is defined as the integral of 
the pseudoscalar glueball correlation function 
\be 
\label{chi_top}
 \chi_{top} = \lim_{V\to\infty} \frac{\langle Q_{top}^2\rangle}{V} 
  = -\int d^4x\; \Pi_P(x) \, , 
\ee
where $Q_{top}$ is the topological charge and $V$ is the volume. 
This integral diverges at short distance, and care has to be taken in 
defining the subtraction scheme. In the present case it is sufficient 
to subtract the contribution from the free gluon bubble, $\Pi_P^0
=6\alpha_s/(\pi^6 x^8)$. This corresponds to subtracting the correlator 
in the infinite (non-cutoff) $AdS_5$ space. The integrand $2\pi^2 
x^2(\Pi_P(x)-\Pi_P^0(x))$ is shown in Fig.~\ref{fig_top_x}. We observe 
that the integrand changes sign, and we find that the topological 
susceptibility vanishes. This result, as well as the subtraction scheme,  
can be checked by computing the topological susceptibility directly 
from the solution of the equation of motion (\ref{eom_a}) for $q^2=0$. 
The only solution that satisfies the boundary conditions $a(0)=1$ 
and $\partial_z a(z_m)=0$ is $a(z)=1$, and the topological 
susceptibility
\be 
\label{chi_5d}
\chi_{top} =-\frac{c^2}{2N_f} 
   \left.\frac{\partial_z a}{z^3} \right|_{\epsilon} .
\ee
vanishes. This result does not depend on the simplifying assumption
$c=const$. It does, however, depend on the choice $\kappa=0$. The
correlation function for $\kappa\neq 0$ is also shown in 
Fig.~\ref{fig_pig}. The anomalous term suppresses the correlator 
and leads to a non-vanishing (and positive) topological susceptibility.
In the simplest model, defined by equ.~(\ref{eom_a}), the topological
susceptibility has a non-perturbative ultra-violet divergence. This 
can be seen by treating the anomaly term as a small correction. In this
limit the topological charge correlator is \cite{Katz:2007tf}
\be
\label{pip_kappa}
\Pi_P(Q)  = -\frac{1}{2N_f}
   \int_0^{z_m} \frac{dz}{z^5} \, \bar{\kappa} \, 
   \left[\frac{1}{2}(Qz)^2 K_2(Qz)\right]^2 ,
\ee
where the expression in the square brackets is the bulk-to-boundary
propagator of the $a$ field in the infinite $AdS$ geometry (for
$\bar\kappa=0$). For $Q^2=0$ the integrand is singular near $z=0$ unless 
$\bar\kappa(z)$ vanishes faster than $z^4$ as $z\to 0$. This problem can 
be understood by comparing equ.~(\ref{pip_kappa}) with the expectation 
from an instanton model, see appendix \ref{sec_app}. In terms of instantons
the integral $dz/z^5$ arises from the integral over the size of the 
instanton, and the power $z^5$ is fixed by classical scale invariance. 
The Bessel function corresponds to the Fourier transform of the 
topological charge density, and the order of the Bessel function is 
determined by classical scale invariance and the dimension of the 
operator $G\tilde{G}$. The instanton contribution in QCD contains 
an extra factor $\exp(-8\pi^2/g^2(z))$, where $8\pi^2/g^2$ is the 
instanton action, and $g(z)$ is the running coupling constant. Asymptotic 
freedom implies that this term scales as $z^b$ (for $z\to0$), where 
$b$ is the first coefficient of the beta function. As a consequence, 
the integral is well behaved for small $z$, but diverges at large
scale sizes. In the holographic model the infrared problem is cured by the 
hard wall cutoff, but the simplest version of the model has an
ultraviolet divergence instead.

\begin{figure}[t]
\begin{center}
\includegraphics[width=11cm]{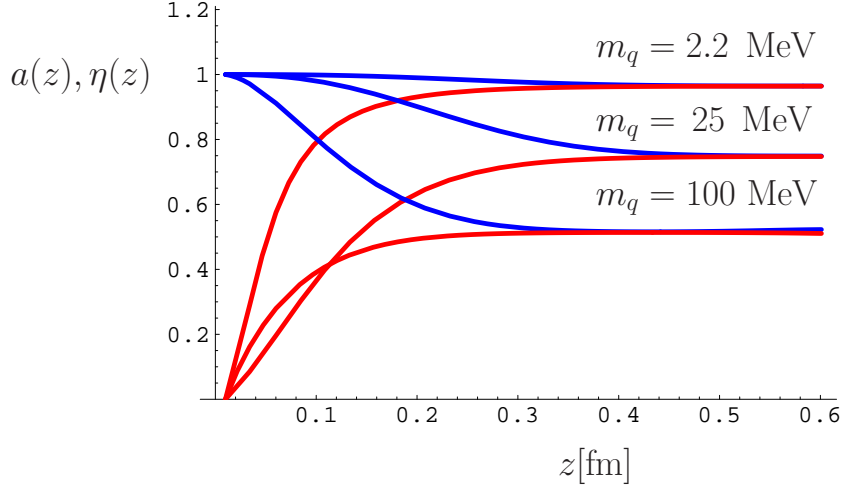}
\end{center}
\caption{\label{fig_top}
Five dimensional $q^2=0$ modes in the pseudoscalar sector. We show 
solutions of the equations of motion for $a(z)$ (blue, solutions start 
at $a(0)=1$) and $\eta^0(z)$ (red, satisfy $\eta^0(0)=0$) for different
values of the quark mass $m$. The topological susceptibility is related
to the negative of the curvature of $a(z)$ near $z=0$. }
\end{figure}

 The problem can be addressed by considering a more complicated 
functional form for $\bar\kappa(z)$. Here, we proceed directly to the 
unquenched theory, which has a finite topological susceptibility.
The topological charge correlator in the theory with three flavors
is shown in Fig.~(\ref{fig_pig}). There is an extra contribution
which is dominated by mixing with the lowest $\eta'$ resonance. 
This term increases the topological charge correlator at long 
distance, and tends to cancel the topological susceptibility 
generated in the pure gauge theory. The topological susceptibility 
can be determined most accurately by solving the equations of motion 
for $Q^2=0$, 
\bea 
\label{top_eom_1}
\partial_z\left( \frac{c^2}{z^3} \partial_z a\right) 
 + \kappa \frac{v^{N_f}}{z^5} \left( \eta^0 -a \right) &=& 0\, , \\[0.2cm]
\label{top_eom_2}
 v^2 \partial_z \eta^0  +  c^2 \partial_z a  &=& 0 \, ,
\eea
subject to the boundary conditions 
$a(0)=1$, $\eta^0(0)=0$, $\partial_z a(z_m)=\partial_z\eta^0(z_m)=0$. 
Solutions to equs.~(\ref{top_eom_1}-\ref{top_eom_2}) for different 
values of the quark masses are shown in Fig.~\ref{fig_top}. We observe 
that the topological susceptibility scales with the quark mass, as 
expected from low energy theorems based on the axial anomaly 
\cite{Witten:1979vv,Veneziano:1979ec}. For $m=2.2$ MeV we obtain 
$\chi_{top}=(70\, {\rm MeV})^3$, in agreement with the expected 
value $\chi_{top}=m\sigma/N_f$. 

\section{Summary}
\label{sec_sum}

 We have studied euclidean coordinate space correlation functions 
in a holographic model of QCD. The correlation functions in the 
flavor non-singlet vector and axialvector channel agree well
(within $\sim 20$\%) with experimental data from hadronic tau
decays. The holographic model does not describe perturbative 
logarithms and power corrections. In principle this can be addressed
by modifying the $AdS_5$ geometry, or by adding higher dimensional 
terms in the five-dimensional action.

 In the pure gauge sector, without an anomaly term, the model 
generates a non-trivial pseudoscalar glueball spectrum but the 
topological susceptibility is zero. The anomaly term gives a 
negative (repulsive) contribution to the pseudoscalar glueball 
correlator and leads to a non-zero topological susceptibility. 
In the full theory an extra attractive contribution arises from
mixing between the pseudoscalar glueball and the pseudoscalar
meson ($\eta'$) field. These two terms cancel as the quark 
mass goes to zero, and the topological susceptibility vanishes. 
The attractive term is longer ranged than the repulsive one, 
so the vanishing of $\chi_{top}$ can be viewed as being due to 
topological charge screening. It also interesting to compare the
results to expectations from the instanton model. The anomaly 
contribution to both the topological charge and the pseudoscalar
meson correlator agree with the structure of the dilute instanton
gas result. 

 In this work we studied a very schematic model based on a slice of 
$AdS_5$. Clearly, it is of interest to study correlation functions in 
``top-down'' models. A very detailed study of anomalous correlators in 
the ${\cal N}=4$ theory was carried out by Dorey et al.~\cite{Dorey:1999pd}
Some attempts to study topology and the $\eta'$ meson in extensions
of the simplest AdS/CFT setup can be found in
\cite{Witten:1998uk,Hashimoto:1998if,Hill:2000rr,Kruczenski:2003uq,Barbon:2004dq,Armoni:2004dc}.
It is also important to further clarify the role of instantons 
in the large $N_c$ limit. Witten argued that instanton effects 
are suppressed in the weak coupling limit of a large $N$ field
theory \cite{Witten:1978bc}. In QCD the situation is not clear,
because the theory is classically conformal (it has instantons
of all sizes), and in the large $N_c$ limit only small instantons
are suppressed. We have previously argued that the instanton size 
distribution in the large $N_c$ limit might be a delta function
\cite{Schafer:2002af}. In the language of holographic QCD this 
corresponds to an anomaly term $\kappa(z)$ which is sharply peaked 
at some critical distance $z^*$ in the fifth dimension. 

Acknowledgments: Part of this work was completed at the Newton Institute 
for Mathematical Sciences during the workshop on Strong Fields, 
Integrability and Strings. I would like to thank the Institute for
their hospitality. This work is supported by a grant from the 
United States Department of Energy, \#DE-FG02-03ER41260. I would 
also like to thank M.~Schwartz for useful correspondence. 

\appendix
\section{Correlation functions in the field of an instanton}
\label{sec_app}

 In this appendix we collect a few results for the single instanton
contribution to hadronic correlation functions. The instanton 
contribution to the pseudoscalar glueball correlation function
is \cite{Geshkenbein:1979vb,Novikov:1979ux,Schafer:1994fd}
\be
\label{pi_inst_g}
\Pi_P(Q)  =-2 \int \frac{d\rho}{\rho^5} \, d(\rho) \, 
 \left[\frac{1}{2}Q^2\rho^2 K_2(\rho Q)\right]^2,
\ee
where the Bessel function $K_2(\rho Q)$ arises from the Fourier 
transform of the topological charge density of an instanton. The
factor 2 arises from adding the contribution of instantons and
anti-instantons. The instanton size distribution is given by 
\cite{'t Hooft:1976fv}
\bea 
\label{d_rho}
 d(\rho) &=& C_{N_c}
    \left(\frac{8\pi^2}{g^2}\right)^{2N_c}
    \exp\left(-\frac{8\pi^2}{g(\rho)^2}\right) 
    \, \prod_f \, m_f^*\rho , \\
m_f^*\rho &=& m\rho-\frac{4\pi^2}{3} \langle \bar{q}q\rangle\rho^3,
  \nonumber \\
C_{N_c} &=&  \frac{0.466\exp(-1.679N_c)1.34^{N_f}}
    {(N_c-1)!(N_c-2)!} , \nonumber 
\eea
where $g(\rho)$ is the one-loop running coupling constant
\be
 \frac{8\pi^2}{g^2(\rho)} = 
    -b\log(\rho\Lambda), \hspace{1cm} 
    b = \frac{11}{3}N_c-\frac{2}{3}N_f.
\ee
The factor $1/\rho^5$ is related to classical scale invariance
and matches the determinant of the metric tensor on $AdS_5$. We 
also note that the factor $(m^*\rho)^{N_f}$ matches the factor 
$v^{N_f}=(mz+\sigma z^3)^{N_f}$ in equ.~(\ref{top_eom_1}). The 
remaining terms in $d(\rho)$ are related to fluctuations around
the classical instanton solution and have no obvious counterpart in 
the holographic model. 

The instanton contribution to the flavor non-singlet pseudoscalar meson 
correlation function is \cite{Shuryak:1982qx}
\be
\label{pi_inst_pi}
\Pi_\pi(Q)  =  2 \int \frac{d\rho}{\rho^5} \, d(\rho) \,
\frac{1}{m^2} \left[ Q\rho K_1(\rho Q)\right]^2.
\ee
This result includes only the contribution of the fermion zero
mode in the quark propagator. The instanton contribution to the 
flavor singlet correlator is the same up to an overall sign.
The Bessel function $K_1(\rho Q)$ is the Fourier transform of
the fermion zero mode. The order is related to the dimension 
of the Fermion field, and matches the bulk-to-boundary propagator
for the pseudoscalar field on $AdS_5$. 

The instanton contribution to the vector current correlation 
function is \cite{Dubovikov:1981bf}
\be 
\label{pi_inst_v}
\Pi_{\mu\nu}(Q) =
  \left(q_\mu q_\nu -\delta_{\mu\nu}q^2\right)
  \frac{4}{3} \int \frac{d\rho}{\rho^5} \, d(\rho) \,
  \left\{ \frac{1}{Q^4} -\frac{3\rho^2}{Q^2}
    \int_0^1 dx \, K_2\left( 2\rho Q/(1-x^2)^{1/2}\right) 
    \right\}.
\ee
The first term is a pure power correction which corresponds 
to the $\langle g^2G^2\rangle$ and $m\langle \bar{q}q\rangle$ 
terms in the OPE. These terms are not present in the holographic 
model. The second term, which is due to non-zero modes of the 
fermion propagator in the instanton field, also has no 
counterpart in the model considered here.


\end{document}